\documentclass[10pt,prb,aps,showpacs,twocolumn]{revtex4}
\usepackage[caption=false]{subfig}

\usepackage{amssymb}
\usepackage{amsmath}
\usepackage{commath}
\usepackage{graphicx,bm}
\usepackage{verbatim}
\usepackage[usenames]{xcolor}
\usepackage{pdfpages}
\usepackage{hyperref}
\usepackage{tikz}

\usepackage[normalem]{ulem} 

\def\be#1\ee{\begin{equation}#1\end{equation}}
\def\ba#1\ea{\begin{align}#1\end{align}}

\newcommand{\reff}[1]{Ref.\,\cite{#1}}
\newcommand{\eqqref}[1]{Eq.\,\eqref{#1}}	
\newcommand{\figgref}[1]{Fig.\,\ref{#1}}

\newcommand{\PRB}[3]{Phys.\,Rev.\,B {\bf #1}, #2 (#3).}
\newcommand{\PRL}[3]{Phys.\,Rev.\,Lett. {\bf #1}, #2 (#3).}

\newcommand{\bxtau} {({\bf x}, \tau)}
\newcommand{\bk}	{{\bf k}}
\newcommand{\bp}	{{\bf p}}

\newcommand{\bx}	{{\bf x}}

\newcommand{\eps}	{\epsilon}

\newcommand{\w}	    {\omega}

\newcommand{\ee}                   {\mathrm{e}}                         

\newcommand{\bb}{{\bf b}}
\newcommand{\mel}{m_{\text{el}}}				
\newcommand{\mph}{m_{\text{ph}}}				

\begin{document}
\title{Large anomalous magnetic moment in three-dimensional Dirac and Weyl semimetals}
\author{E.C.I. van der Wurff}
\email{e.c.i.vanderwurff@uu.nl}
\author{H.T.C. Stoof}
\affiliation{Institute for Theoretical Physics and Center for Extreme Matter and Emergent Phenomena, Utrecht University,  Princetonplein 5, 3584 CC Utrecht, The Netherlands}
\date{\today}
\begin{abstract}
We investigate the effect of Coulomb interactions on the electromagnetic response of three-dimensional Dirac and Weyl semimetals. In a calculation reminiscent of Schwinger's seminal work on quantum electrodynamics, we find three physically distinct effects for the anomalous magnetic moment of the relativistic-like quasiparticles in the semimetal. In the case of non-zero doping, the anomalous magnetic moment is finite at long wavelengths and typically orders of magnitude larger than Schwinger's result. We also find interesting effects of one of the three new Hamiltonian terms on the topological surface states at the interface between vacuum and a Weyl semimetal. We conclude that observation of these effects should be within experimental reach.
\end{abstract}

\maketitle
\section{Introduction}
Two breakthroughs in the history of quantum mechanics were the discovery of spin by Stern and Gerlach in 1922 and the formulation of the Schr\"odinger equation in 1926 \cite{Gerlach,Schroedinger}. These two concepts were combined in 1927, when Pauli generalized Schr\"odinger's equation to take into account the Zeeman interaction of an electron's spin with an external magnetic field, assuming the value $g_{m} = 2$ for its dimensionless magnetic moment \cite{Pauli1}. This theory was subsequently put on a firm footing by Dirac in 1928 when he derived the fully relativistic wave equation that now carries his name \cite{Dirac}. A natural result in his derivation of Pauli's theory was a dimensionless magnetic moment of precisely $g_{m}=2$ for the electron. However, in 1947 a slight deviation from $g_{m}=2$ was measured \cite{RabiZacharias}. It took only a year before this discrepancy was resolved, when Schwinger used the powerful tools of the newly developed quantum field theory to calculate the anomalous magnetic moment of the electron due to its coupling to the photon \cite{Schwinger}.

One important consequence of the anomalous magnetic moment becomes clear when it is added to the Dirac equation for an electron in an external magnetic field. In its absence, the energy spectrum consists of Landau levels that are all doubly degenerate, except for the spin-polarized zero-energy Landau level. However, this spin degeneracy is removed when an anomalous magnetic moment is present \cite{Incera}. This spin splitting is reminiscent of the situation that occurs when spin degeneracy is removed in a Dirac semimetal by breaking time-reversal or inversion symmetry, which splits the Dirac cone into two Weyl cones. The resulting material is nowadays referred to as a Weyl semimetal.

Recently, Dirac and Weyl physics has been discovered in materials such as Cd$_3$As$_2$ and Na$_3$Bi, and TaAs, respectively \cite{Liu1,Neupane,Cava,Liu2,Jeon,Savrasov}. The last material has also been proposed to exhibit anomalous transport properties due to its topological nature such as an anomalous Hall effect and a chiral anomaly \cite{Burkov1,Burkov2,Franz,Adler,Jackiw}. In view of these rapid developments it is of interest to take a closer look at the anomalous magnetic moment of the electrons in these materials. However, there is now a fundamental difference with Schwinger's calculation: he considered the case of \emph{massive} electrons in the vacuum, whereas the electrons in a Weyl semimetal are described by \emph{massless} electrons around the band-touching points. In the case of massless electrons in a vacuum, one would find that the anomalous magnetic moment contains an infrared divergence. Luckily, there is a way around this conundrum. Due to their coupling to a non-zero density of electrons, photons acquire an effective mass at long wavelengths. This results in a screened Coulomb potential between the electrons in the material, providing an infrared cutoff that renders the anomalous magnetic moment finite.

There have been previous investigations into the role of \emph{unscreened} Coulomb interactions between the electrons in a Weyl semimetal, including the effect on charge transport \cite{Vishwanath}, the Berry curvature \cite{Abanin} and the thermodynamic stability \cite{Nomura}. Additionally, vertex corrections were investigated both numerically \cite{Gonzalez1} and using the renormalization group \cite{Gonzalez2}. Here we instead include the effects of screening and provide an analytical calculation of the anomalous magnetic moment resulting from the transversal part of the vertex correction.

We start our discussion by deriving the screened Coulomb potential in doped Weyl semimetals in Sec.\,\ref{sec:screening}. Subsequently, we investigate how the coupling of the electrons to an external electromagnetic field is changed by including these screened electron-electron interactions. By calculating the so-called transverse vertex corrections in Sec.\ref{sec:vertexcorrection} we find that the anomalous magnetic moment of a Weyl fermion generates three distinct Hamiltonian terms, one of which constitutes a Rashba spin-orbit coupling, whereas the other two form a Zeeman-like effect. Due to the large ratio of the free electron mass and the effective mass of the photon, both effects are non-negligible. To illustrate that these effects can have observable consequences, we consider the influence of the anomalous Rashba spin-orbit coupling on the topological edge states of a Weyl semimetal in Sec.\,\ref{sec:surfacestates}. Finally, we discuss our results in Sec.\,\ref{sec:discussion}.
\section{Screening in Dirac and Weyl semimetals} \label{sec:screening}
We start by considering a three-dimensional semimetal with Dirac dispersion $\eps(k) = \hbar v_F k$, in terms of its Fermi velocity $v_F$, which we always assume to be much less then the speed of light $c$ so that we are allowed to neglect retardation effects and current-current interactions. Moreover, throughout the following we will ignore logarithmic interaction corrections. For instance, the Fermi velocity will be taken constant, independent of the doping of the material. To preserve generality we consider a system with $g$ degenerate Dirac cones that split up into $g/2$ pairs of Weyl cones when time-reversal symmetry is broken. In principle, each of these pairs of cones can be separated in momentum space with its own time-reversal symmetry breaking vector ${\bf b}$ that acts as an internal Zeeman field.

The dielectric function for such a material has been studied extensively and is known analytically in the random-phase approximation \cite{DasSarma1, DasSarma2, Xiao, Pesin, Agarwal}. This is a good approximation because the number of Weyl nodes $g$ is typically large and higher-order corrections are surpressed in powers of $1/g$ \cite{DasSarma3}. In terms of its complex frequency $z$ and wavenumber $q$ the dielectric function equals $\eps(q,z) = 1 - V(q)\Pi(q,z)$, with the Coulomb potential $V(q) = e^2/\eps_0 q^2$. The polarizability at zero temperature can be written as
\ba
&\Pi(q,z) = \frac{-g}{24\pi^2 \hbar v_F}\bigg[8k_F^2 + q^2\log\bigg(\frac{v_F^2\Lambda^2}{v_F^2q^2 - z^2} \bigg)         \nonumber \\
&\phantom{=}- q^2\sum_{\sigma,\sigma' = \pm}G\bigg( \frac{\sigma'z+2v_Fk_F}{\sigma v_Fq} \bigg)H\bigg(\frac{\sigma' z - \sigma v_F q}{2v_Fk_F}\bigg) \bigg],
\ea
where $k_F$ is the Fermi wavenumber and we defined the dimensionless functions $H(x) \equiv \ln(1+1/x)$ and $G(x) \equiv \big( x^3 - 3x + 2 \big)/4$. Here, $\Lambda$ is an ultraviolet cutoff scale. Its physical origin lies in the fact that in a real material, the linear Dirac dispersion is only an approximation that holds up to a certain energy scale. 

The screened Coulomb potential follows from the static retarded dielectric function $\eps(q,0^+)$ and reads, again up to logarithmic corrections,
\be \label{eq:coulomb}
V_{\text{sc}}(q) = \frac{V(q)}{\eps(q,0^+)} \text{ }\overset{q\rightarrow0}{=}\text{ } \frac{e^2}{\eps_0}\frac{1}{q^2 + \xi^{-2}},
\ee
with the screening length $\xi \equiv \sqrt{\pi/2 g \bar{\alpha}k_F^2}$ in terms of the dimensionless effective fine-structure constant $\bar{\alpha} \equiv c \alpha/v_F = e^2/4 \pi\eps_0 \hbar v_F$. This also defines an effective photon mass $m_{\text{ph}} \equiv \hbar/\xi c$.

We note explicitly that, due to $v_F/c \ll 1$, this screened Coulomb potential does not depend on the separation in momentum space between a pair of Weyl cones. Furthermore, the screened Coulomb potential reduces to an ordinary one in the case of zero doping. Additionally, the numerically transformed Coulomb potential shows very small-amplitude Friedel oscillations at large distances, but as the exponential Yukawa-decay is dominant, we will neglect those here \cite{Agarwal}. We also noted above that the effective photon mass in the screened Coulomb potential will act as an infrared cutoff. For the typical parameters $k_F = 0.04$ \AA$^{-1}$, $g=12$ and $v_F = c/300$ \cite{Cava, Savrasov}, we have $k_F\xi \approx 0.24 $ and a photon mass which is three orders of magnitude smaller than the free electron mass $\mel$. This implies that the Coulomb potential has a relatively large range and as a result vertex corrections can be of significant size, as we will show now.
\section{Transversal vertex corrections in Weyl semimetals} \label{sec:vertexcorrection}
We specialize to a Weyl semimetal with broken time-reversal symmetry characterized by a shift of $2\bf b$ between a pair of Weyl points in momentum space. Although we shall ultimately consider the zero-temperature limit, we set up our framework at a non-zero temperature $T$. The Weyl fermions are described by the spinor $\psi(\bx,\tau)$. Thus, the imaginary-time action for a single Weyl cone with chemical potential $\mu$ and chirality $\eps = \pm$ then reads
\be \label{eq:action}
S_0 = \hbar\int d\tau d\bx
\psi^{\dagger}(\bx,\tau)\bigg[\partial_{\tau} - \frac{\mu}{\hbar} - v_F{\bm \sigma}\cdot\big(i \epsilon \nabla + {\bf b}\big) \bigg]\psi\bxtau,
\ee
where $\tau \in [0,\hbar\beta]$ with $\beta = (k_B T)^{-1}$ and $\bm{\sigma}$ is the vector of Pauli matrices. The emergent Lorentz symmetry allows us to use a relativistic notation with unit metric. Using this notation, we couple the Weyl fermions with charge $-e$ to the external four-potential $A^{\mu} = (\phi/v_F,{\bf A})$ via the minimal substitution $\hbar\partial_{\tau} \rightarrow \hbar\partial_{\tau} - e\phi(\bx,\tau)$ and $-i\hbar\nabla \rightarrow -i\hbar\nabla + e {\bf A}(\bx,\tau)$. This way the coupling terms in the action can be written as $-ev_F A_{\mu}\gamma^{\mu}$, with $\gamma^{\mu} = (1, -\eps\sigma)$ the bare vertex. The final ingredient that we add to our theory is interactions between the fermions via the screened Coulomb potential $V_{\text{sc}}(q)$ presented in \eqqref{eq:coulomb}.
\begin{figure}[t!]
\includegraphics[scale=.7]{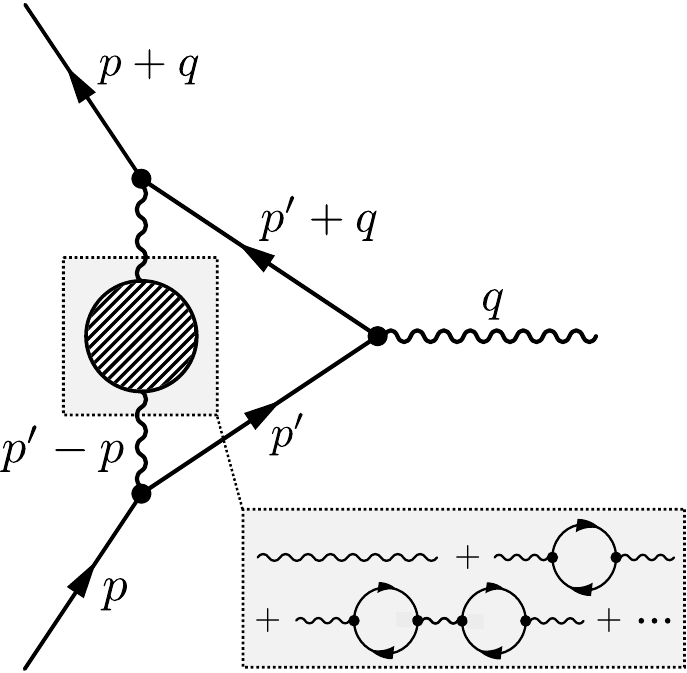} 
\caption{Feynman diagram for the vertex correction. Photons are denoted by whiggly lines and fermions by straight lines. The dashed circle and inset indicate that we are using the screened Coulomb interaction in the random-phase approximation.} \label{fig:feynman}
\end{figure}

We next perform perturbation theory in the bare vertex and the screened Coulomb interactions. We absorb the location of the Weyl cone into the wavenumber by $\bm{\sigma}\cdot(\eps\bk-{\bf b}) \equiv \eps\bm{\sigma}\cdot \bp$. Then the bare Matsubara propagator of the fermion is $G_0(p) = (i\w_n + \mu/\hbar- \eps v_F \bm{\sigma}\cdot\bp)^{-1}$ and we find that the lowest-order vertex correction for a system of volume $V$ reads
\be \label{eq:vertex}
\hbar\beta V \Gamma^{\mu}(p,q) = -\sum_{\bp', n'}G_0(p' + q)\gamma^{\mu} V_{\text{sc}}(\bp' - \bp) G_0(p'),
\ee
where $p^{\mu}$ and $q^{\mu}$ are the incoming fermion and photon wavenumber. The vertex correction is depicted graphically in terms of a Feynman diagram in \figgref{fig:feynman}. 

The total vertex needs to preserve gauge invariance, which can be checked with the so-called Ward-Takahashi identity. For the bare vertex it is given by the relation $v_F q_{\mu}\gamma^{\mu} = G_0^{-1}(q+p) - G_0^{-1}(p)$, while for the vertex correction, the Ward-Takahashi identity reads \cite{Schroeder}
\be
v_F q_{\mu}\Gamma^{\mu}(p,q) = -\Sigma(q + p) + \Sigma(p),
\ee
where $\hbar\beta V\Sigma (p) = -\sum_{\bp',n'}G_0(p')V_{\text{sc}}(\bp' - \bp)$ is the self-energy in the so-called $G_0W$-approximation.

The vertex correction in \eqqref{eq:vertex} has longitudinal and transversal parts. The longitudinal part is divergent in the ultraviolet and upon regularization it ultimately leads to a renormalization of the Fermi velocity as a function of the doping. This has been investigated thoroughly in \reff{DasSarma4}. Since the renormalization is only logarithmic in $\mu$, it is for our purposes allowed to not consider it here.
Instead, we aim to investigate the transversal part of the vertex correction. Besides using the Ward-Takahashi identity, we fix the decomposition of the vertex correction in its transversal and longitudinal parts uniquely by demanding that both the longitudinal and the transversal vertex are hermitian. To make analytic progress we specialize to zero temperature and perform perturbation theory in the external photon momentum up to first order. This is valid if we consider external electric and magnetic fields that vary slowly in space and time. Finally, gauge invariance enures that everything can be written in terms of the external electric and magnetic fields $\bf E$ and $\bf B$. We thus arrive at a correction to the action in \eqqref{eq:action} that reads
\ba \label{eq:vertexfinal}
-e v_F A_{\mu} \Gamma^{\mu}_{\text{transv.}}(\bp) &= -\big[\bm{\mu}_1(\bp)\times\bm{\sigma}\big]\cdot\frac{{\bf E}}{v_F} \nonumber \\
&\phantom{=}- \big[ \bm{\mu}_2(p) - \eps {\bm{\mu}}_1(\bp) \big] \cdot {\bf B},
\ea
in terms of the magnetic moments $\bm{\mu}_1(\bp) \equiv \mu_1(p)\bp$ and $\bm{\mu}_2(p) \equiv \mu_2(p)\bm{\sigma}$. The first term constitutes a Rashba-spin-orbit coupling, while the second and third term together form a Zeeman-like effect.
Note that the third term is proportional to $\eps$, indicating that its sign depends on the chirality of the Weyl cone under consideration.

Upon defining the function $f(x) \equiv \ln\big[1 + (k_F\xi + x)^2\big] - \ln\big[1 + (k_F\xi - x)^2\big]$, the magnitudes $\mu_i(p)$, scaled by the Bohr magneton $\mu_B = e \hbar / 2 \mel$, are given by
\ba \label{eq:mu1}
\frac{\mu_1(p)}{\mu_B}&= \frac{ec\alpha}{4\pi \mu_B \xi p^3}
\bigg\{
4p\xi - 2\text{arctan}\big( \xi [k_F + p]\big) \nonumber \\
&\phantom{=}+  2\text{arctan}\big(\xi [k_F - p]\big) - k_F\xi f(p\xi)
\bigg\} \nonumber \\
&\overset{p\rightarrow0}{=} \frac{\alpha}{2\pi} \frac{\mel}{\mph} \bigg(\frac{2[1 - (k_F\xi)^2]}{3[1+(k_F\xi)^2]^2}\bigg)\xi,
\ea
and
\begin{figure}[t!]
\includegraphics[trim={0cm 0cm 0cm 0cm},clip,scale=.8]{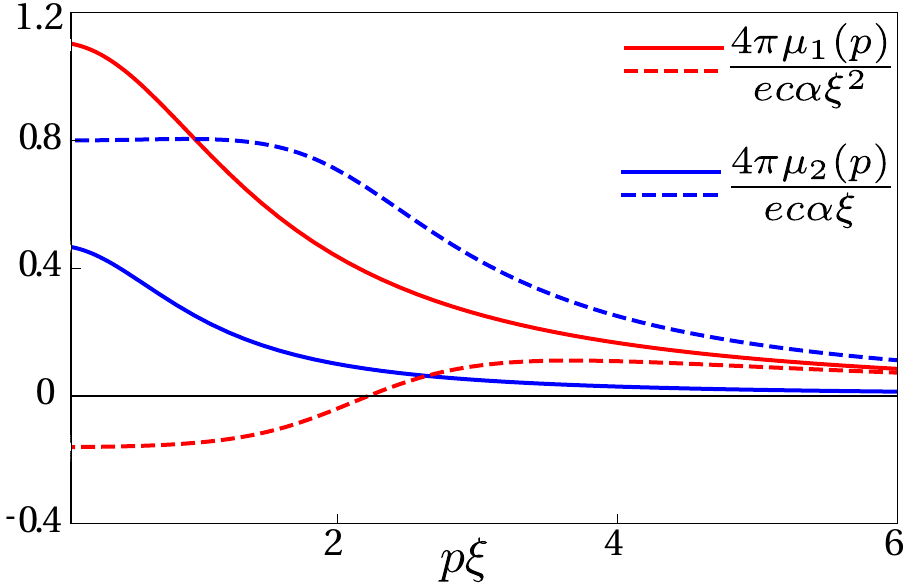} 
\caption{(color online). Plot of dimensionless magnetic moments for $k_F\xi = 1/4$ (solid lines) and $k_F\xi=2$ (dashed lines). Note that $\mu_1(p)$ changes sign for small $p$ when $k_F\xi<1$.} \label{fig:moments}
\end{figure}
\be \label{eq:mu2}
\frac{\mu_2(p)}{\mu_B} = \frac{ec\alpha f(p\xi)}{8 \pi \mu_B p} \text{ }\overset{p\rightarrow0}{=}\text{ } \frac{\alpha}{2\pi} \frac{\mel}{\mph}\bigg(\frac{2k_F\xi}{1+(k_F\xi)^2}\bigg),
\ee
where $\alpha/2\pi$ is Schwinger's result in terms of the fine-structure constant $\alpha$ and $\mel/\mph\gg1$, as discussed previously. Equations \eqref{eq:vertexfinal}, \eqref{eq:mu1} and \eqref{eq:mu2} form the central result of this paper. We stress that although $\alpha/2\pi \approx 1.2\cdot10^{-3}$ is small, it is multiplied by the large fraction $m_{\text{el}}/m_{\text{ph}} \approx 1.6\cdot10^3$, indicating that our results are between two and three orders of magnitude larger than those of Schwinger. Indeed, we find for $k_F\xi\approx0.24$ in the long-wavelength limit $\mu_1k_F/\mu_B \approx 0.25 $ and $\mu_2/\mu_B \approx 0.85$. Hence, these anomalous effects are substantial.

One interesting property is that both the magnetic moment $\bm{\mu}_1$ and the linear combination $\bm{\mu}_2 - \eps {\bm{\mu}}_1$ can change sign for small $p$ and $k_F\xi<1$. As $k_F\xi = \sqrt{\pi/2g\bar{\alpha}}$, this means that the strength of the effective fine-structure constant determines the sign of the Zeeman or Rashba effect. We have plotted the two moments for two values of $k_F\xi$ in \figgref{fig:moments}.
\begin{figure*}
\includegraphics[scale=.65]{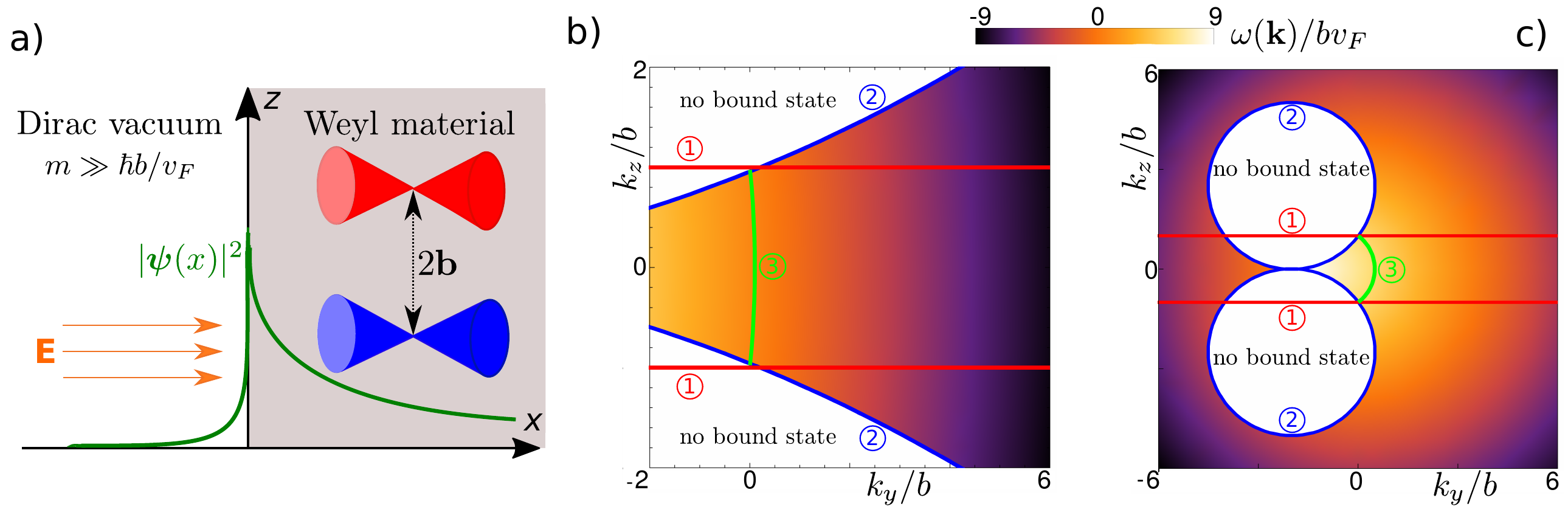}
\caption{(color online). a) Schematic illustration of the proposed experimental set-up. In this case the electric field is perpendicular to the surface. b) Density plot of the dimensionless dispersion $\omega_{\bk}/bv_F$ as a function of $k_y/b$ and $k_z/b$ for $\bar{E} = 1/10$. c) Same as in b), for $\bar{E}=1/2$. The white regions cannot support a bound state. In both figures the red lines [and \textcircled{1}] indicate the boundary of the area where there are bound states for zero electric field, wheres the blue lines [and \textcircled{2}] do the same in the case of a non-zero electric field. The green lines [and \textcircled{3}] indicate the Fermi arc for a non-zero electric field.} \label{fig:boundstates}
\end{figure*}

Yet another important property is that the two magnetic moments are finite in the long-wavelength limit. The crucial ingredient for obtaining this finite long-wavelength behavior is a non-zero doping. This causes the Coulomb potential to be screened, yielding an infrared cutoff for the loop-integral in the vertex correction. Indeed, in the limit $k_F \rightarrow 0$, or $\xi\rightarrow\infty$, we obtain the infrared divergence $\mu_i(p)\sim p^{-2}$. Exactly the same behavior was recently found in a holographic calculation at the critical Dirac point \cite{Jacobs}.
\section{Rashba-modified topological surface states}  \label{sec:surfacestates}
To consider observable effects of the anomalous magnetic moment, we now investigate how the Rashba spin-orbit coupling due to a constant external electric field $\bf E$ influences the topological surface states of a Weyl semimetal. The effect of the Zeeman-like anomalous terms will be discussed in future work. We would like to stress that the Rashba spin-orbit coupling we derived in \eqqref{eq:vertexfinal} does not follow from a Peierls substitution in the Weyl Hamiltonian. Its true origin lies in the Coulomb interactions between the electrons.

We consider a simple model system inspired by Refs.\,\cite{Tewari,Zyuzin}. We have an interface at $x=0$ of a Dirac vacuum with mass $m$ for $x<0$ and a massless Weyl semimetal with a time-reversal symmetry breaking vector ${\bf b} = b \hat{z}$ for $x>0$, with $b>0$. The mass $m\gg \hbar b/v_F$ acts as the equivalent of a work function for the electrons to leave the Weyl semimetal into the Dirac vacuum. Such a system has a surface state localized around $x=0$ that propagates in the $yz$-plane. This state is chiral because it only admits the dispersion $\omega_{\bk} = -v_Fk_y$, whereas the dispersion with the other sign is not a solution. Additionally, it only exists for $|k_z| < b$, which leads to the so-called Fermi arc of the system \cite{Savrasov}. A schematic image of this set-up is presented in \figgref{fig:boundstates}(a).

Now we ask how these properties change when we turn on a constant external electric field which induces the anomalous Rashba spin-orbit coupling. For $x>0$ the Hamiltonian is block diagonal with its constituent blocks $H_{\pm}$ corresponding to the different chiralities $\eps = \pm$. In the long-wavelength limit it reads
\be
\frac{H_{\pm}(\bk)}{\hbar v_F} = \pm \big[\bk \mp {\bf b}\big]\cdot \bm{\sigma} - \big[\bar{\bf E}\times(\bk \mp {\bf b})\big]\cdot\bm{\sigma},
\ee
where we introduced the dimensionless electric field $\bar{\bf E} \equiv \mu_1(0){\bf E}/\hbar v_F^2$ and we reintroduced the time-reversal symmetry breaking vector by setting $\bp = \bk \mp \bb$. The wavenumbers along the $y$ and $z$-direction are good quantum numbers, such that the eigenvalue problem becomes one-dimensional in the $x$-coordinate 

We proceed by solving the eigenvalue problems of the Dirac and Weyl Hamiltonian for $x<0$ and $x>0$ separately and demand continuity of the spinors at $x=0$. This yields a four-by-four matching matrix, which should have a vanishing determinant to support a bound state located around $x=0$. There are three physically distinct situations. An electric field in the $y$-direction yields the same dispersion relation and Fermi arc as in the case of zero electric field. In the case of an electric field in the $z$-direction, we find a modified dispersion $\omega_{\bk} = - v_F\sqrt{1+\bar{E}^2}k_y$, while the Fermi arc is still given by $|k_z|<b$. Finally, the most interesting situation is the one where the electric field is in the $x$-direction, perpendicular to the surface. In this case we find a dispersion relation that reads
\be \label{eq:dispersion}
\omega_{\bk} = -v_F\frac{(1-\bar{E}^2)bk_y + \bar{E}(k_y^2 + k_z^2 - b^2)}{\sqrt{(b+\bar{E}k_y)^2 + \bar{E}^2k_z^2}},
\ee
which only is a solution in the exterior of the two circles defined by $\big[k_y + \frac{b}{\bar{E}} \big]^2 + \big[k_z \pm \frac{b}{2}\big( 1 + \frac{1}{\bar{E}^2} \big) \big]^2  = \frac{b^2}{4}\big[ 1+ \frac{1}{\bar{E}^2} \big]^2$. Hence, there are no bound states with wavenumbers in the interior of these circles because the one-dimensional problem then corresponds to a topologically trivial band insulator. This is drastically different from the condition $|k_z|>b$ for zero electric field. Furthermore, by demanding $\omega_{\bk}=v_Fk_F$ we find that the Fermi arc is a part of a circle that increases in size for larger values of the electric field. This should be contrasted to the case of zero electric field, where the Fermi arc is just a straight line. We show a density plot of the dispersion relation and Fermi arc for two values of $\bar{E}$ in Fig.\,\ref{fig:boundstates}(b) and (c). 

This dramatic change in the Fermi arc when an external electric field is applied could be an experimental signature to detect the anomalous Rashba spin-orbit coupling. Additionally, we deduce from Figs.\,\ref{fig:boundstates}(b) and (c) that there are bound states with a certain wavenumber which seize to be a bound state when a critical value for the externally applied electric field is reached. This could be another experimental signature of the presence of the anomalous Rashba spin-orbit coupling. To make quantitative predictions for a specific material, one needs to consider a more detailed model of the surface. However, we stress that the qualitative behavior obtained above is determined by topology and will remain the same.
\section{Conclusion and discussion} \label{sec:discussion}
In conclusion, we have calculated the first-order vertex correction for a three-dimensional Weyl semimetal in which the massless electrons interact via a screened Coulomb potential. We have shown that the correction is orders of magnitude larger than Schwinger's result for the anomalous magnetic moment for massive electrons in quantum electrodynamics. Finally, we have demonstrated that the anomalous Rashba spin-orbit coupling has an observable effect on the surface states located between a Weyl semimetal and vacuum, as well as on the corresponding Fermi arcs. A surface Rashba effect due to the inevitable breaking of inversion symmetry at the surface of the material\cite{Bluegel} does not modify these latter observations. Indeed, the mere existence of the surface states is determined by the topological nature of the band
structure in the bulk, which is characterized by the bulk Hamiltonian that includes the Rashba-like spin-orbit coupling coming from the transversal vertex correction. Hence, a Rashba effect on the surface changes the dispersion relation in \eqqref{eq:dispersion}, but will not change the boundaries of the regions in the $(k_y,k_z)$-plane that determine whether or not there is a surface state.

In future work we plan to investigate the influence of the Zeeman-like term on the bound surface states of a Weyl semimetal. It would also be interesting to explore if the anomalous effects derived in this paper leave a distinct signature in the quasiparticle interference pattern of a Weyl semimetal \cite{Bernevig,Fritz,Hasan2}.
\newline

It is our pleasure to thank Guido van Miert, Vivian Jacobs, Panos Betzios, Umut G\"ursoy and Lars Fritz for useful discussions and the latter also for reading the manuscript. This work is supported by the Stichting voor Fundamenteel Onderzoek der Materie (FOM) and is part of the D-ITP consortium, a program of the Netherlands Organisation for Scientific Research (NWO) that is funded by the Dutch Ministry of Education, Culture and Science (OCW).

\onecolumngrid

\end{document}